\documentstyle[prl,twocolumn,aps,epsfig,floats]{revtex}
\begin{document}

\draft

\title{Phase Separation in the Two-Dimensional 
Systems of Strongly Correlated Electrons;
the Role of Spin Singlet Pairs 
on Hole Pairing Contribution to Hole-rich Phase
}
\author{Sung-Sik Lee, Sul-Ah Ahn, and Sung-Ho Suck Salk$^{1,2}$}
\address{$^{1}$ Department of Physics, 
Pohang University of Science and Technology,
Pohang 790-784, Korea\\
$^{2}$ Korea Institute of Advanced Study, Seoul 130-012, Korea} 
\date{\today}
\maketitle
 
\begin{abstract}
By paying attention to the hole-doped two-dimensional systems 
of antiferromagnetically (strongly) correlated electrons, 
we discuss the cause of hole-rich phase formation in association with 
phase separation. 
We show that the phase diagram obtained from the Maxwell's construction 
in the plane of temperature vs. hole density is consistent with
one derived from the evaluation
of hole-rich and electron-rich phases in real space. 
We observe that the formation of a hole-rich phase 
is attributed to the aggregation of hole pairs induced by
spin singlet pairs present in the pseudogap phase 
and that a direct involvement of correlations between hole pairs
are not essential for phase separation.
\end{abstract}
 

Currently there has been an activated interest 
in the study of stripe modulations
owing to strong experimental evidences observed 
\cite{Zimmermann,Tranquada,Shen,Jorgensen,Hundley,Hammel,Weidinger,Harshman,Cho,Borsa,Dong,Hunt,Nikolaev}
involving hole-doped high $T_c$ cuprates.
Phase separation, a phenomenon of broken symmetry
at low energies can be manifested as a stripe phase 
or a phase-separated inhomogeneity.
At a critical temperature, the phase separation appears 
owing to the divergence of compressibility 
 or the zero inverse compressibility.
The onset of phase separation can be obtained from the evaluation of 
the zero inverse compressibility  
using the well-known Maxwell's construction.
Thus far, various numerical approaches
\cite
{Zaanen,EmeryKivelsonLin,Seibold,Luchini,White,Putikka,Dagotto,Prelovsek,Castellani,Poliblanc,Kohno,Hellberg,Shraiman,Calendra,Pryadko,Cosentini,Gimm,WhiteScalapino,HellbergManousakis}
to the Hubbard and $t-J$ Hamiltonians
have successfully revealed the existence of phase separation
in the hole-doped two dimensional systems of strongly correlated electrons.
They include the Hatree-Fock (HF), 
the Lanczos exact diagonalization (LED), 
the density matrix renormalization group (DMRG), 
a fixed node quantum Monte Carlo (FNQMC) and 
Green's function Monte Carlo (GFMC) methods.
Most recently a slave-boson functional integral (SBFI) method 
of Gimm and Salk\cite{Gimm} has been proposed
to reveal a good agreement 
 with earlier exact numerical studies\cite{EmeryKivelsonLin,Hellberg} 
of phase separation boundary. 
However it remains to be seen
whether the phase separation obtained from the Maxwell's construction 
can, indeed, show phase-separated distributions 
of charge and spin in real space.
Currently there exists a great controversy 
\cite{WhiteScalapino,HellbergManousakis,Prelovsek,Hellberg}
over the issue of whether the phase separation 
accompanied by the hole-rich phase 
arises as a consequence of correlations between pairs of holes or not.
Here we address this issue, based on 
our slave-boson functional integral approach\cite{Gimm,SSLee},
and take the advantage of its usefulness to study 
the role of charge and spin degrees of freedom 
on phase separation.

The $t-J$ Hamiltonian to deal with the hole doped systems 
of antiferromagnetically correlated electrons is written,
\begin{eqnarray}
H&=&-t\sum_{\langle i,j\rangle\sigma}\left(c_{i\sigma}^{\dagger}c_{j\sigma} 
+\mbox{H.c.}\right)
+J\sum_{\langle i,j\rangle}\left({\bf S}_i\cdot{\bf S}_j 
- \frac{n_in_j}{4}\right)\\
\label{tj_ham_1}
&=& -t\sum_{\langle i,j\rangle\sigma} \left(c_{i\sigma}^{\dagger}c_{j\sigma} 
+\mbox{H.c.}\right) \nonumber\\
&&-\frac{J}{2} \sum_{\langle i,j\rangle}
\left(c_{i\uparrow}^{\dagger}c_{j\downarrow}^{\dagger}
-c_{i\downarrow}^{\dagger}c_{j\uparrow}^{\dagger}\right)
\left(c_{j\downarrow}c_{i\uparrow}-c_{j\uparrow}c_{i\downarrow}\right).
\label{tj_ham_2}
\end{eqnarray}
In order to explicitly see physics involved with 
the spin and charge degrees of freedom,
we rewrite the above equation in the $U(1)$ slave-boson representation 
with the use of $c_{i\sigma}=f_{i\sigma}b_i^{\dagger}$ 
as a composite of spinon $f_{i\sigma}$ and holon $b_i$ operators
subject to the single occupancy constraint 
$b_i^{\dagger}b_i + \sum_{\sigma}f_{i\sigma}^{\dagger}f_{i\sigma}=1$
\begin{eqnarray}
H_{t-J}&=&-t\sum_{\langle i,j\rangle\sigma}f_i^{\dagger}f_j b_j^{\dagger}b_i
-\frac{J}{2} \sum_{\langle i,j\rangle} b_ib_j b_j^{\dagger}b_i^{\dagger} \nonumber\\
&&\times\left(f_{i\uparrow}^{\dagger}f_{j\downarrow}^{\dagger}
-f_{i\downarrow}^{\dagger}f_{j\uparrow}^{\dagger}\right)
\left(f_{j\downarrow}f_{i\uparrow}-f_{j\uparrow}f_{i\downarrow}\right).
\label{tj_slave_boson}
\end{eqnarray}
The above $U(1)$ slave-boson expression can also be deduced 
from our earlier $SU(2)$ slave-boson representation\cite{SSLee} 
of the $t-J$ Hamiltonian.
The second term shows spinon pairing (spin singlet) 
interactions between adjacent sites. 
Both terms explicitly manifest coupling 
between the charge (holon) and spin (spinon) degrees of freedom.

\begin{figure}[h]
\centering
        \epsfxsize = 7cm
        \epsfysize = 6cm
        \epsffile{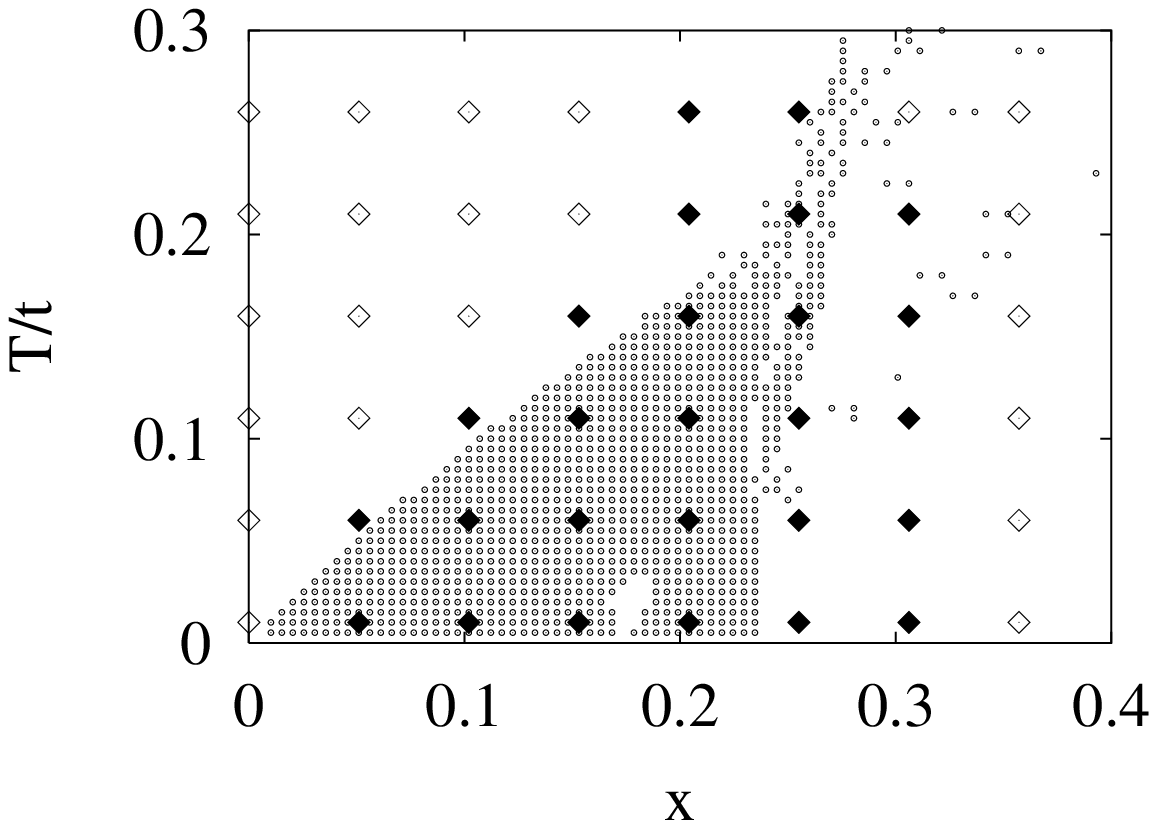}
\vspace*{5mm}
        \caption{}
\label{phase_U4}
\end{figure}

We attempt a multi-faceted study of phase separation 
by employing various levels of approaches;
the Hatree-Fock (HF)\cite{SSLee}
of Hubbard Hamiltonian for a qualitative, comparative study of phase separations, 
the Monte Carlo diagonalization (MCD)\cite{RaedtLinden} for 
an accurate study of 
 correlated electron systems
and the slave-boson functional integral (SBFI) method
for an investigation of the role of charge (hole) and spin degrees of freedom 
on phase separation.
Extending our earlier HF approach\cite{SSLee}
of Hubbard Hamiltonian to a study of temperature dependence 
of charge and spin distributions in real space\cite{SSLeePreprint},
we first examine the domain of phase separation
in the plane of temperature vs. hole density
and compare the phase separation domain derived from the Maxwell's construction 
with the one obtained from the direct (real) space calculations
of electron-rich and hole-rich phases. 
Based on HF calculations 
with $U=4t$, in Fig. \ref{phase_U4}
we display the predicted phase separation diagram in the plane of  
 the temperature $T$ vs. hole density $x$ with the two different approaches;
one from the use of the Maxwell's construction 
(denoted by shaded area in Fig. \ref{phase_U4})  
and the other from the direct evaluations 
of hole (charge) and electron (spin) distributions 
(denoted as solid diamonds) in the real space of $14\times 14$ lattice.
Temperature dependent HF total energies for Maxwell's construction 
are obtained from the solution of the following self-consistent equations 
(Eq. (\ref{self_consistent_m_mu})) 
involving the quasi-particle energy, 
\begin{equation}
E_{\bf k}^{\pm}= \frac{U}{2}(1-x) - \mu \pm \epsilon_{\bf k}
\label{self_consistency}
\end{equation}
with $\epsilon_{\bf k} = \sqrt{[-2t(\cos k_x + \cos
  k_y)]^2 + \left( \frac{m U}{2} \right)^2 }$ and 
 $m=|\langle c_{i \uparrow}^{\dagger}c_{i \uparrow}
-c_{i \downarrow}^{\dagger}c_{i \downarrow}\rangle|$, 
by determining the uniform staggered magnetization $m$ 
and the chemical potential $\mu$
\begin{eqnarray}
1 & = & \frac{1}{N} \sum^{'}_{\bf k} \frac{U}{2 \epsilon_{\bf k}} 
\left[ \tanh \left( \frac{E_{\bf k}^+}{2T} \right) - \tanh \left(
    \frac{E_{\bf k}^-}{2T} \right) \right] \nonumber \\
x & = & \frac{1}{N} \sum^{'}_{\bf k} \left[
\tanh \left( \frac{E_{\bf k}^+}{2T} \right) + 
\tanh \left( \frac{E_{\bf k}^-}{2T} \right) \right].
\label{self_consistent_m_mu}
\end{eqnarray}
Encouragingly the two results showed a large overlapping domain 
of phase separation in the $T-x$ plane.
The present mean field results are at best qualitative.

\begin{figure}[h]
\centering
        \epsfxsize = 6cm
        \epsfysize = 6cm
        \epsffile{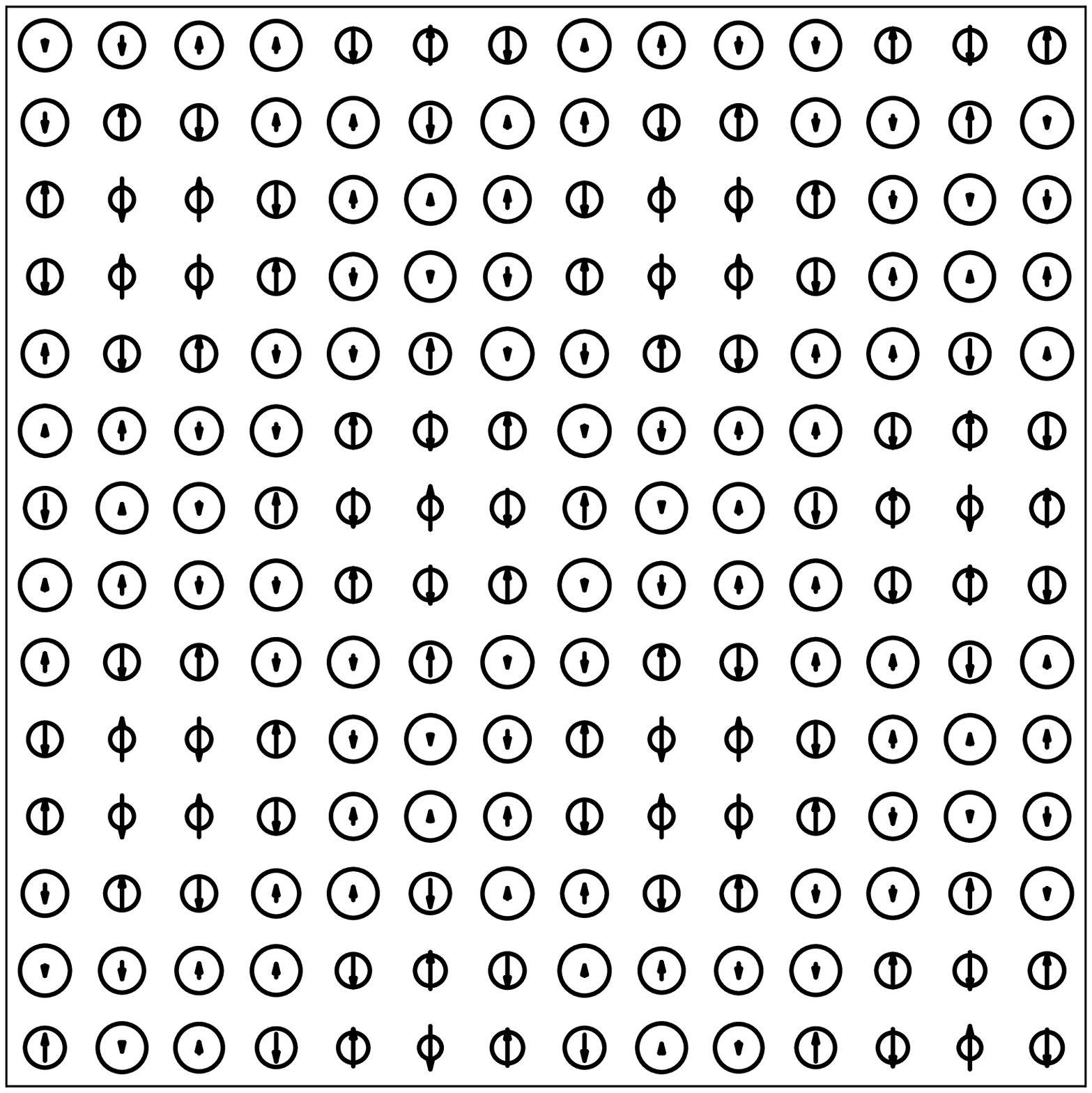}
        \vspace*{5mm}
        \caption{}
\label{gp_T10_d50}
\end{figure}

In order to show the doping dependence of phase separation in real space, 
we now examine the electron and hole distributions  
 interior and exterior 
to the phase separation boundary.
At a chosen hole density of $x = 0.25$ 
near but below a critical value $x_c\simeq 0.3$ at $T=0.01t$,
a stripe phase no longer appears and, 
instead, an inhomogeneous phase separation is revealed,
as is shown in Fig. \ref{gp_T10_d50}.
Although not shown here, such inhomogeneous structures 
are prevalent near but below the critical hole density $x_c$,  
but they eventually disappear to yield uniform phase structures
beyond $x_c$.
It should be noted that despite the fact that no correlations exist
between the Hatree-Fock quasiparticles,
the phase separation is seen to occur as well known.
They are not accurate for 
strongly correlated (large $U$) electron systems. 
However, it is to be noted that with currently available, 
numerically exact methods, the present level of real space calculations 
of temperature dependent phase separation for a large square lattice 
are not easily feasible.
It is of great interest to see whether there exists a possibility 
of 'metastable' phase-separation in a certain region of temperature and doping 
where density fluctuations are excessively large 
as is shown by the results from the Maxwell's construction.
This case happened near $T=0.01t$ and $x=0.25$, 
showing a metastable region of phase separation. 
In this small domain, the density fluctuations were predicted to be large.
A detailed study for verification is necessary in the future.

\begin{figure}[h]
\centering
\epsfig{file=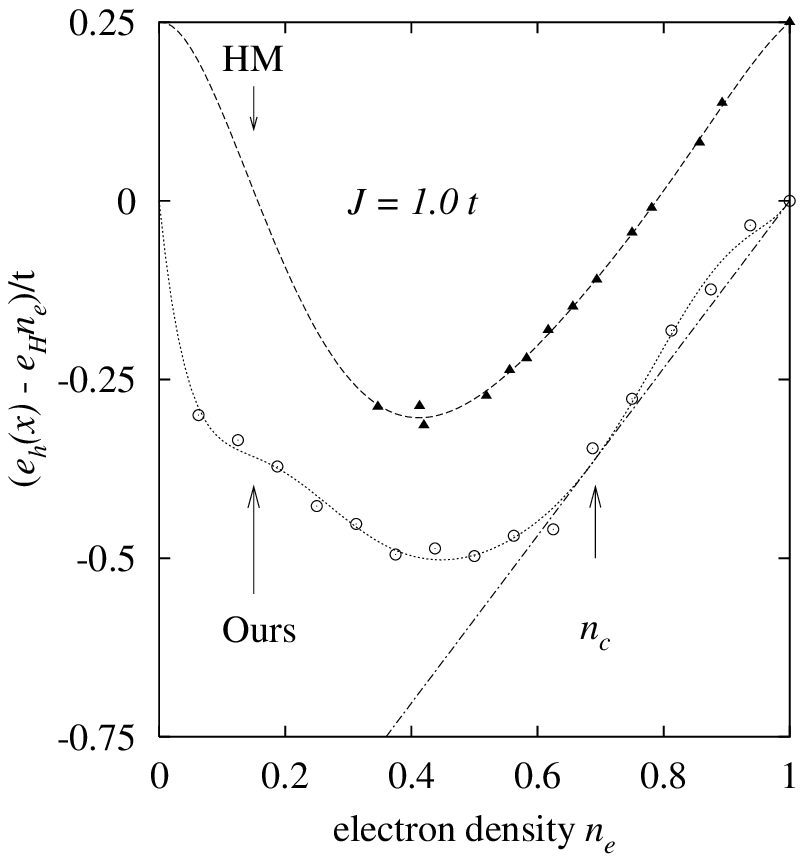, width=7cm}
\vspace*{5mm}
\caption{}
\label{maxwell_j1}
\end{figure}

\begin{figure}[h]
\centering
\epsfig{file=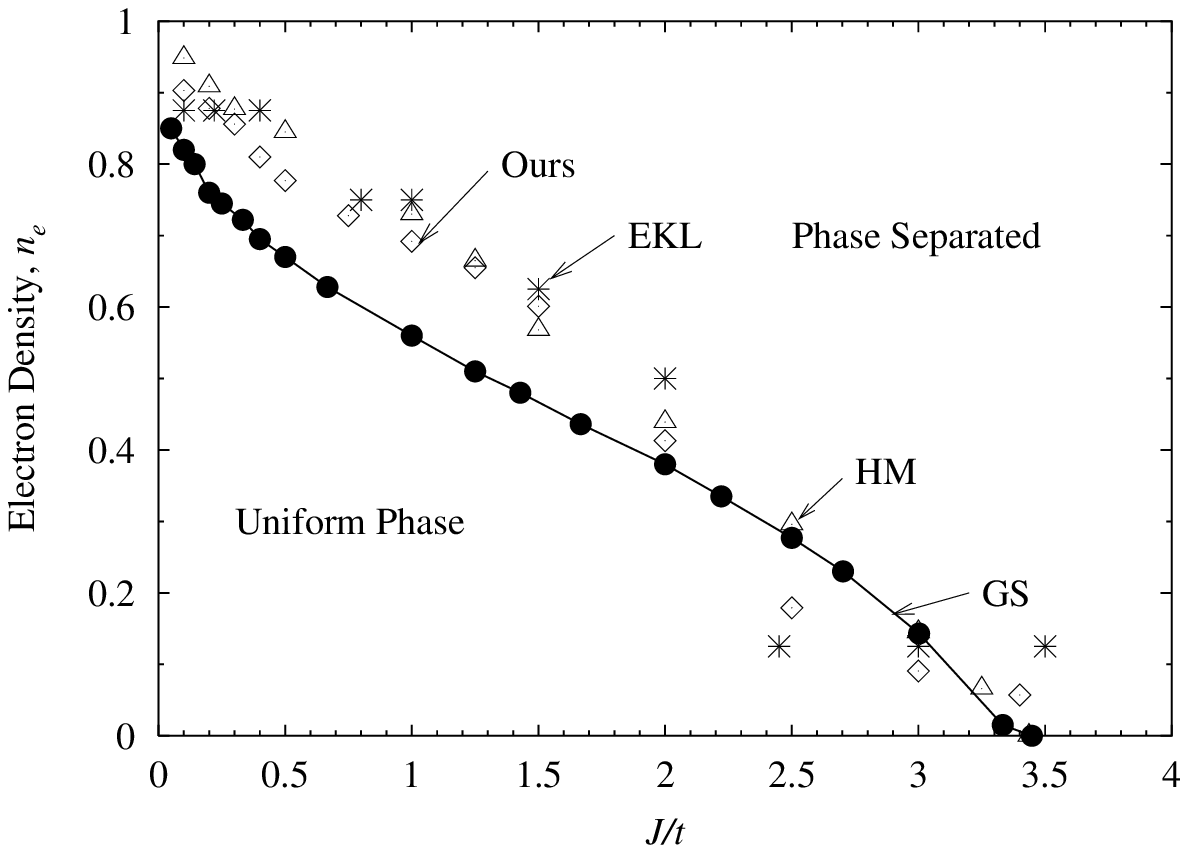, width=7cm}
\vspace*{5mm}
\caption{}
\label{various_ps}
\end{figure}

For an accurate account of strongly correlated electrons 
for the study of phase separation 
as a function of antiferromagnetic coupling strength $J$,
we now take
the MCD method\cite{RaedtLinden} 
of the $t-J$ Hamiltonian, 
Eq. (1) with $J=1.0t$ for a $4\times 4$ square lattice. 
Fig. \ref{maxwell_j1} shows the computed ground state energy 
(in units of $t$) shifted by a linear factor $e_H n_e$
as a function of the electron density $n_e$ ($n_e = 1-x$). 
The solid triangle represents 
the GFMC calculation by Hellberg and Manousakis\cite{Hellberg}$^{c}$
and the open circles, our present results.
From the Maxwell's construction the critical electron density (hole density)
 is predicted to be $n_c=0.684$ ($x_c=0.316$) 
compared to the value of $n_c = 0.730$ ($x_c=0.270$) obtained from 
the GFMC.
The overall variation of curvature as a function of electron (hole) density 
is grossly similar between the two approaches,
although the predicted ground state energies are not the same.
In Fig. \ref{various_ps}, we display our MCD predicted phase separation boundary 
in the plane of electron (hole) density vs. antiferromagnetic interaction strength 
(Heisenberg coupling constant) in units of $t$
and make comparison with other methods.
The MCD agreed very well with the GFMC of Hellberg and Manousakis
\cite{Hellberg}$^{c}$, 
the LED of Emery {\em et al.}
\cite{EmeryKivelsonLin} 
and the SBFI ($U(1)$ slave-boson functional integral) result 
of Gimm and Salk\cite{Gimm}. 
A salient feature is that all of these methods  
yielded a similar phase separation boundary,
by showing a smoothly decreasing (increasing) trend  
of the critical electron (hole) density with an intercept near $J/t \sim 3.5$
as the antiferromagnetic interaction strength $J/t$ increases.  

However, the accurate MCD calculations 
can not readily resolve the current controversy over the issue of 
whether the effect of correlations between the hole pairs 
is the primary cause of forming the hole-rich phase.
For this cause we now explore 
the $U(1)$ slave-boson functional integral approach 
of the $t-J$ Hamiltonian (Eq. (\ref{tj_slave_boson})).
This method is advantageous to examine the role of 
the charge and spin degrees of freedom 
or the role of the holon and spinon degrees of freedom 
on phase separation.
Earlier we reported that 
this method\cite{Gimm} also showed a satisfactory phase separation boundary 
for all $J$ values 
($J/t \leq 1$ and $J/t > 1$)
in general agreement with other numerical studies
\cite{EmeryKivelsonLin,Hellberg}. 
The spin (spinon) degrees of freedom shown in the second term 
in Eq. (\ref{tj_slave_boson}) allows 
spinon pairing (spin singlet formation) interactions between adjacent sites.  
Indeed it has been shown from the $U(1)$ slave-boson theories
that the spinon pairing, that is, the spin singlet pair order 
appears below the pseudogap (spin gap) temperature $T^*$
\cite{KotliarLiu,SSLee}.
This indicates that the motion of paired holes rather than
the independent motion of separated holes is energetically preferred 
in the presence of surrounding electron spin singlet pairs (spinon pairs) 
below the pseudogap (spin gap) temperature $T^*$.
This is because the hole pairs (holon pairs) can readily migrate 
to the occupied sites of spin pairs (spinon pairs) 
involving no spin-bond breaking.  

In summary, we showed that 
the phase separation diagram obtained from Maxwell's construction 
in the plane of temperature vs. hole density is consistent with 
one derived from the real space (direct lattice) calculations 
of hole-rich and electron-rich phases; 
 inhomogeneous phase separation  
appears near $x_c$ at finite temperatures;
 the formation of the hole-rich phase for phase separation is attributed 
to the aggregation of hole pairs induced by 
spin singlet pairs which exist in the spin-gap phase. 

One of us (S.H.S.S.) greatly acknowledges generous supports 
of POSRIB (2001) project at Pohang University of Science and Technology 
and Korean Ministry of Education (Hakjin Excellence Leadership Program 2001).
We thank Chan-Ho Yang for computational assistances.

\newpage
\centerline{FIGURE CAPTIONS}
\begin{itemize}

\item[FIG. 1]
The temperature (in the unit of hopping energy $t$) dependence 
of phase separation as a function of hole doping density $x$
based on the Hatree-Fock approximation with $U=4t$ 
for a $14\times 14$ square lattice.
The shaded region represents phase separation obtained 
from the Maxwell's construction; 
the solid diamond, the phase separation obtained 
from the numerical evaluation of charge and spin distributions
in real space and the open diamond, the uniform phase.

\item[FIG. 2]
Spin (electron) and charge (hole) distribution at 
$x  = 50/196 (\simeq 0.25)$  $T=0.01t$.
The large circles represent holes (charges) 
and smaller ones, electrons (spins).
The size of circles represents the degree of hole-richness
or electron-richness. 

\item[FIG. 3]
Maxwell's construction (dotted line) from 
predicted ground state energies denoted by open circles;
the energies (in units of $t$) are 
shifted by a linear factor $-e_H n_e$,
$\left(e_h(x)-e_H n_e \right)/t$ as a function of electron density for $J=1.0t$
with a $4\times 4$ lattice.
For comparison the Green's function Monte Carlo result\cite{Hellberg}$^{d}$
 (denoted by HM) with various lattice sizes up to $11\times 11$ is displayed.

\item[FIG. 4]
Phase separation for the hole-doped systems
of antiferromagnetically correlated electrons 
in the plane of Heisenberg coupling strength, $J/t$ 
and the electron density, $n_e=1-x$. 
The triangles denoted by HM represent
the Green's function Monte Carlo prediction
of Hellberg and Manousakis 
\cite{Hellberg}$^{c}$
 and the stars denoted by EKL,
the exact diagonalization result of Emery {\em et al.}\cite{EmeryKivelsonLin}.
The solid circles represent the results 
from the $U(1)$ slave-boson functional integral approach
of Gimm and Salk\cite{Gimm} and 
diamonds, our present MCD results.

\end{itemize}

\begin{thebibliography}{99}
%
\bibitem{Zimmermann} 
M. v. Zimmermann, A. Vigliante, T. Niem\"oller, N. Ichikawa, T. Frello, 
J. Madsen, P. Wochner, S. Uchida, N. H. Andersen, J. M. Tranquada, D. Gibbs,
 and J. R. Schneider,  
Europhys. Lett. {\bf 41}, 629 (1998).
\bibitem{Tranquada} 
J. M. Tranquada, J. D. Axe, N. Ichikawa, A. R. Moodenbaugh, Y. Nakamura, 
and S. Uchida, 
Phys. Rev. Lett. {\bf 78}, 338 (1997);
J. M. Tranquada, J. D. Axe, N. Ichikawa, Y. Nakamura, S. Uchida, 
and B. Nachumi, 
Phys. Rev. B. {\bf 54}, 7489 (1996);
J. M. Tranquada, B. J. Sternlieb, J. D. Axe, Y. Nakamura, and S. Uchida, 
 Nature {\bf 375}, 561 (1995);
P. Wochner, J. M. Tranquada,  D. J. Buttrey, and V. Sachan, 
Phys. Rev. B, {\bf 57}, 1066 (1998);
J. M. Tranquada, D. J. Buttrey, V. Sachan, and J. E. Lorenzo, 
Phys. Rev. Lett. {\bf 73}, 1003 (1994); 
V. Sachan, D. J. Buttrey, J. M. Tranquada, J. E. Lorenzo, and G. Shirane, 
Phys. Rev. B. {\bf 51}, 12 742 (1995); 
J. M. Tranquada, J. E. Lorenzo, D. J. Buttrey, and V. Sachan, {\em ibid.}
{\bf 52}, 3581 (1995).
%
\bibitem{Shen} 
A. Lanzara, P. V. Bogdanov, X. J. Zhou, S. A. Kellar, D. L. Feng, 
E. D. Lu, T. Yoshida, H. Elsaki, A. Fusimori, K. Kishio, 
J.-I. Shhimoyama, T. Noda, S. Uchida, Z. Hussain, and Z. X. Shen, 
Nature {\bf 412}, 510 (2001); references therein.
%
\bibitem{Jorgensen} 
J. D. Jorgensen, B. Dabrowski, S. Pei, D. G. Hinks, L. Soderholm, 
B. morosin, J. E. Schriber, E. L. Venturini, and D. S. Ginley,
Phys. Rev. B {\bf 38}, 11 337 (1988).
%
\bibitem{Hundley}
M. F. Hundley, J. D. Thompson, S-W. Cheong, Z. Fisk, and J. E. Schirber,
Phys. Rev. B {\bf 41}, 4062 (1990).
%
\bibitem{Hammel}
S.-H. Lee and S-W. Cheong, Phys. Rev. lett. {\bf 79}, 2514 (1997);
P. C. Hammel, A. P. Reyes Z. Fisk, M. Takigawa, J. D. Thompson,
R. H. Heffner, S-W. Cheong, and J. E. Schirber, 
Phys. Rev. B, {\bf 42}, 6781 (1990);
P. C. Hammel {\em et al.}, Physica C {\bf 185-189}, 1095 (1991).
%
\bibitem{Weidinger}
A. Weidinger, Ch. Niedermayer, A. Golnik, R. Simon, and E. Recknagel, 
Phys. Rev. Lett. {\bf 62}, 102 (1989).
%
\bibitem{Harshman}
D. R. Harshman, G. Aeppli, B. Batlogg, G. P. Espinosa, R. J. Cava, 
A. S. Cooper, L. W. Rupp, E. J. Ansaldo, and D. Ll. Williams,
Phy. Rev. Lett. {\bf 63}, 1187 (1989).
\bibitem{Cho}
J. H. Cho, F. Borsa, D. C. Johnston, and D. R. Torgeson, 
Phys. Rev. B {\bf 46}, 3179 (1992);
J. H. Cho, F. C. Chou, and D. C. Johnston,
Phys. Rev. Lett. {\bf 70}, 222 (1993).
%
\bibitem{Borsa}
F. Borsa, P. Carretta, A. Lascialfari, D. R. Torgeson, F. C. Chou, 
and D. C. Johnston, 
Physica C {\bf 235-240}, 1713 (1994);
F. Borsa, P. Carretta, J. H. Cho, F. C. Chou, Q. Hu, D. C. Johnston, 
A. Lascialfari, D. R. Torgeson, R. J. Gooding, N. M. Salem, and K. J. E. Vos, 
Phys. Rev. B {\bf 52}, 7334 (1995);
F. C. Chou and D. C. Johnston, Phys. Rev. B {\bf 54}, 572 (1996).
%
\bibitem{Dong}
X. L. Dong, Z. F. Dong, B. R. Zhao, Z. X. Zhao, X. F. Duan, L.-M. Peng, 
W. W. Huang, B. Xu, Y. Z. Zhang, S. Q. Guo, L. H. Zhao, and L. Li, 
Phys. Rev. Lett. {\bf 80}, 2701 (1998).
%
\bibitem{Hunt}
A. W. Hunt, P. M. Singer, K. R. Thurber, and T. Imai, 
Phys. Rev. Lett. {\bf 82}, 4300 (1999); references therein.;
%
\bibitem{Nikolaev}
E. G. Nikolaev, H. B. Brom, and A. A. Zakharov, 
Phys. Rev. B {\bf 62}, 3050 (2000).
%
\bibitem{Zaanen} 
J. Zaanen and O. Gunnarsson, Phys. Rev. B {\bf 40}, 7391 (1989);
J. Zaanen and P. B. Littlewood, Phys. Rev. B. {\bf 50}, 7222 (1994).
%
\bibitem{EmeryKivelsonLin} 
V. J. Emery, S. A. Kivelson, and H. Q. Lin, 
Phys. Rev. Lett. {\bf 64}, 475 (1990);
S. A. Kivelson, V. J. Emery, and H. Q. Lin, 
Phys. Rev. B {\bf 42}, 65223 (1990);
S. A. Kivelson and V. J. Emery, 
in {\em Strongly Correlated Electronic Matrials: 
the Los Alamos Symposium, 1993},
edited by K. S. Bedel {\em et al.}
(Addison-Wesley, Reading, MA, 1994);
V. J. Emery and S. A. Kivelson, Physica {\bf 209C}, 597 (1993).
%
\bibitem{Seibold}
G. Seibold, C. Castellani, C. Di Castro, and M. Grilli, 
Phys. Rev. B {\bf 58}, 13 506 (1998).
%
\bibitem{Luchini} 
M. U. Luchini, M. U. Ogata, W. O. Puttika, and T. M. Rice,
Physica {\bf 185-189C}, 141 (1991);
W. O. Putikka, M. U. Luchini, and T. M. Rice, 
Phys. Rev. Lett. {\bf 68}, 538 (1992).   
%
\bibitem{White} 
S. R. White, Phys. Rev. Lett. {\bf 69}, 2863 (1992);
Phys. Rev. B {\bf 48}, 10 345 (1993).
%
\bibitem{Putikka} 
W. O. Putikka, M. U. Luchini, and T. M. Rice, 
Phys. Rev. Lett. {\bf 68}, 538 (1992).
%
\bibitem{Dagotto} 
E. Dagotto, Rev. Mod. Phys. {\bf 66}, 763 (1994);
H. Fehske, V. Waas, H. R\"oder, and H. B\"utter, Phys. Rev. B {\bf 44}, 8473 (1991);
D. poilblanc, {\em ibid.} {\bf 52}, 9201 (1995).
%
\bibitem{Prelovsek} 
P. Prelov\v{s}ek and X. Zotos, Phys. Rev. B {\bf 47}, 5984 (1993).
%
\bibitem{Castellani} 
C. Castellani, C. Di Castro, and M. Grilli, Phys. Rev. B {\bf 40}, 7391 (1989).
%
\bibitem{Poliblanc} D. Poliblanc, Phys. Rev. B {\bf 52}, 9201 (1995).
%
\bibitem{Kohno} M. Kohno, Phys. Rev. B {\bf 55}, 1435 (1997).
%
\bibitem{Hellberg} 
a) C. S. Hellberg and E. Manousakis, Phys. Rev. Lett. {\bf 78}, 4609 (1997);
b) J. Phys. Chem. Solids {\bf 59}, 1818 (1998);
c) Phys. Rev. B {\bf 61}, 11 787 (2000).
%
\bibitem{Shraiman} 
B. Shraiman and E. Siggia, Phys. Rev. Lett. {\bf 60}, 740 (1988);
{\bf 61}, 467 (1988); 
M. Boninsegni and E. Manousakis, Phys. Rev. B {\bf 45}, 4877 (1992);
{\bf 46}, 560 (1992).
%
\bibitem{Calendra} 
M. Calendra, F. Becca, and S. Sorella, Phys. Rev. Lett. {\bf 81}, 5185 (1998). 
%
\bibitem{Pryadko} 
L. P. Pryadko, S. Kivelson, and D. W. Hone, Phys. Rev. Lett. {\bf 80}, 5651 (1998).
%
\bibitem{Cosentini}
A. C. Cosentini, M. Capone, L. Guidoni and G. B. Bachelet, 
Phys. Rev. B {\bf 58}, R 14685 (1998).

\bibitem{Gimm} 
T. H. Gimm and Sung-Ho Suck Salk, Phys. Rev. B {\bf 62}, 13 930 (2000). 
%
\bibitem{WhiteScalapino} 
S. R. White and D. J. Scalapino, Phys. Rev. Lett. {\bf 84}, 3021 (2001); 
{\bf 80}, 1272 (1998);{\bf 81}, 3227 (1998); 
Phys. Rev. B {\bf 60}, 753 (1999).
%
\bibitem{HellbergManousakis} 
S. Hellberg and E. Manousakis, Phys. Rev. lett. {\bf 84}, 3022 (2001);
{\bf 83}, 132 (1999).
%
\bibitem{SSLee} 
S.-S. Lee and Sung-Ho Suck Salk, Phys. Rev. B {\bf 64}, 052501 (2001);
T.-H. Gimm, S.-S. Lee, S.-P. Hong and Sung-Ho Suck Salk, Phys. Rev. B
{\bf 60}, 6324 (1999).
%
\bibitem{RaedtLinden} 
H. De Raedt and W. von der Linden, Int. J. Mod. Phys. C {\bf 3}, 97 (1992);
Phys. Rev. B {\bf 45}, 8787 (1992);
W. von der Linden, Phys. Rep. {\bf 220}, 53;
H. De Raedt and M. Frick, Phys. Rep. {\bf 231}, 107 (1993);
%
\bibitem{SSLeePreprint} 
S.-P. Hong, S.-S. Lee, and Sung-Ho Suck Salk, Phys. Rev. B {\bf 62}, 
14880 (2000).
%
\bibitem{KotliarLiu} 
G. Kotliar and J. Liu, Phys. Rev. B {\bf 38}, 5142 (1988);
M. U. Ubbens and P. A. Lee, Phys. Rev. B {\bf 46}, 8434 (1992);
 {\bf 49}, 6853 (1994); 
X. G. Wen and P. A. Lee, Phys. Rev. Lett. {\bf 76}, 503 (1996);
{\bf 80}, 2193 (1998);
N. Nagaosa and P. A. Lee, Phys. Rev. B {\bf 61}, 9166 (2000).
\end{thebibliography}
\end{document}